\documentclass[cits]{PoS}

\title{Exploring Partially Confined Phases}

\ShortTitle{Exploring Partially Confined Phases}

\author{\speaker{Michael C. Ogilvie}%
         \thanks{MCO and JCM gratefully acknowledge the support of the U.S. Dept of Energy.}\\
        Washington University\\
        E-mail: \email{mco@physics.wustl.edu}}

\author{Peter N. Meisinger\\
        Washington University\\
        E-mail: \email{pnm@physics.wustl.edu}}

\author{Joyce C. Myers\\
        Washington University\\
        E-mail: \email{jcmyers@wustl.edu}}

\abstract{Phases of SU(N) gauge theories in which the global Z(N) symmetry breaks spontaneously to a subgroup Z(L) can be realized by adding appropriate Wilson line terms to the gauge action. These phases are partially confining, in the sense that quarks are confined but bound states of L quarks are not. At temperatures large compared to the normal deconfinement temperature, the phase diagram, pressure, string tensions, and 't Hooft loop surface tensions can be calculated analytically. Approximate scaling laws emerge naturally for both string tensions and surface tensions.}

\FullConference{The XXV International Symposium on Lattice Field Theory\\
		 July 30 - August 4 2007\\
		 Regensburg, Germany}

\begin{document}

\section{Introduction}

In recent work on $SU(3)$ and $SU(4)$ gauge theories, Myers and Ogilvie have shown that the addition of a term of the form $T\lambda_A Tr_A P$ to the gauge theory Lagrangian, where $P$ is the Polyakov loop, produces novel phases at temperatures above the pure gauge deconfinement 
transition temperature \cite{Myers:2007vc}.  The new phase of $SU(4)$ is a partially confined phase, where the $Z(4)$ global symmetry spontaneously breaks to $Z(2)$, indicating
confinement of quarks but not diquark states.
The new phase of $SU(3)$ is distinct from the confined and deconfined phases,
and occurs at intermediate values of $\lambda_A$. For large values of $\lambda_A$, the confined phase of $SU(3)$ is restored. 
In both $SU(3)$ and $SU(4)$, a theoretical analysis of the phase structure
and thermodynamics is in good agreement with lattice simulation results.
This has prompted us to a detailed analytical study of the general case of $SU(N)$.
For a class of models with extended actions, we can study many new 
high-temperature phases, including confining and partially confining
phases. We add to the gauge theory Lagrangian an external potential
of the form
\begin{equation}
V_{ext}=\sum_{k=1}^{[N/2]}\lambda_k Tr_F P^k Tr_F P^{\dagger k}.
\end{equation}
which is both gauge-invariant and center-symmetric, {\it i.e.}, invariant
under global $Z(N)$ symmetry. We are able to calculate
 the phase structure, thermodynamics, string tensions, 't Hooft loop surface tensions, and Debye screening masses associated with
all these phases at high temperatures

Our work is partially motivated  by
our previous work on an effective potential
approach to the deconfinement transition 
\cite{Meisinger:2002kg,Meisinger:2005ws,Meisinger:2006hg}.
Additional motivation comes from the work of
Davies {\it et al.}  \cite{Davies:1999uw,Davies:2000nw}
on the role of monopoles in supersymmetric
theories on $R^{3}\times S^{1}$,
and of Diakonov {\it et al.} \cite{Diakonov:2004jn}
on calorons in $SU(N)$ gauge
theories at finite temperature.
These works
have demonstrated 
how topological
excitations can make non-perturbative
contributions to the effective potential
that are crucial in determining phase
structure and associated properties.
In both cases, the leading contribution of the topological excitations
to the effective potential is a term of the form 
$\lambda_{1}Tr_F P \,Tr_F P^{\dagger}$
with $\lambda_1$ positive.
On the other hand, normal particles, including the
gauge bosons themselves, make a negative contribution
to $\lambda_1$ \cite{Meisinger:2001fi}.
Generally, a negative value of $\lambda_1$ favors the 
$Z(N)$-breaking deconfined
phase. However, positive $\lambda_1$  favors 
$Tr_{F}P=0$, a defining property of the confined phase.
For $N>3$, the additional terms in $V_{ext}$
are needed to give
confinement for all non-zero $N$-ality states.

\section{Form of the Effective Action}

The form of the effective potential
for Polyakov loop eigenvalues is known
from perturbation theory for weak coupling
\cite{Gross:1980br,Weiss:1980rj},
and should be valid for temperatures
far above the deconfinement temperature
of the pure gauge theory.
This perturbative effective action gives us considerable information
about high-temperature behavior, including
the pressure and the Debye screening mass in the plasma.
Kink solutions give analytical predictions
for 't Hooft loop surface tensions $\rho_k$
 \cite{Bhattacharya:1990hk,Bhattacharya:1992qb}
that can be compared with lattice results 
\cite{Bursa:2005yv,de Forcrand:2005rg}.

At the level of approximation at which we are working,
the addition of $V_{ext}$ to the Lagrangian
also adds it to the effective potential.
We thus consider an effective action
for Polyakov loops of the form
\begin{equation}
S_{eff}=\int d^3x\, 
\frac{T}{g^{2}}\left(\nabla\theta_{k}\right)^{2}-\frac{2T^{3}}{\pi^{2}}\sum_{n=1}^{\infty}\frac{1}{n^{4}}
\left(Tr_{F}P^{n}Tr_{F}P^{\dagger n}-1\right)
+\sum_{k=1}^{\left[N/2\right]}\lambda_{k}TrP^{k}TrP^{\dagger k}
\end{equation}
where the $\theta$'s are eigenvalues of $P$. In a gauge where $A_0$ is time-independent and diagonal we have
$P_{jk}=\delta_{jk}\exp\left[i\theta_j\right]$.
This expression contains a kinetic term obtained by a reduction of the gauge field action as well as an effective potential term obtained from a one-loop 
calculation of the free energy of the gauge bosons in the presence
of a background Polyakov loop. To the standard form of the high-temperature
effective action, we have added $V_{ext}$, which is
gauge-invariant and center-symmetric.
If all the $\lambda_k$'s are positive, the minimum
of $V_{ext}$ occurs for a unique set of
Polyakov loop eigenvalues that are permuted
by a $Z(N)$ symmetry transformation \cite{Meisinger:2002kg}.
If we denote the corresponding Polyakov loop
for an $SU(N)$ model by $P_{0N}$,
then the
$Z(N)$ symmetry transformation $P_{0N}\rightarrow zP_{0N}$
is equivalent to a gauge transformation $gP_{0N}g^\dagger$.
This condition leads immediately to a set of conditions
$Tr_F P^k_{0N}= 0$ for $k=1$ to $N-1$ consistent
with $Z(N)$ symmetry. 
Thus if $\lambda_k \gg T^3$ for all $k$ at high temperatures,
we can recover the confined phase in a regime where
$g(T)$ is small, and perturbation theory is presumably valid.
It is only necessary to
include terms up to $k=[N/2]$
because the string tensions in different
$N$-alities obey $\sigma_{N-k}=\sigma_{k}$.
In contrast, 
the one-loop perturbative potential terms are minimized by breaking $Z(N)$ completely, with minima for $P\in Z(N)$,
corresponding to the deconfined phase.
It is the interplay of the different potential terms
that leads to a remarkably rich phase structure.

\section{Order Parameters and Phase Diagram for $SU(4)$ and $SU(6)$}

We begin by considering the possible phases of the gauge group $SU(4)$.
The center of $SU(4)$ is $Z(4)$, and there is only one non-trivial sub-group, $Z(2)$.
A phase in which $Z(4)$ breaks to $Z(2)$ can be characterized by
$\left< Tr_F P \right> = 0$ and $\left< Tr_F P^2 \right> \neq 0$.
The other possible phases are the  deconfined phase,
characterized by
$\left< Tr_F P \right>\neq 0$ and $\left<  Tr_F P^2\right> \neq 0$,
and the confined phase, which has
$\left< Tr_F P \right> = 0$ and $\left<  Tr_F P^2 \right> = 0$.
All of the phase transitions
between different phases will involve a jump
in one or more order parameters, and are
first order.

The simplest case to analyze occurs when
$| \lambda_k | \gg T^3$ for all $k$.
In this case, the symmetry-breaking
solutions will generally be maximal in the sense
that the magnitude $|Tr_F P^k|$ will be $N$,
unless it is zero. 
The phase structure for $SU(4)$, obtained by
minimizing
$V_{ext}$
is shown in Fig. 1.
The effect of $O(T^3)$ corrections is to
shift the location of the tricritical point away from
the origin, and accordingly shift the critical lines
both horizontally and vertically.
This effect is clearly seen in
 the $SU(4)$ simulations of \cite{Myers:2007vc}.
 Even though $\lambda_2 = 0$, the $Z(2)$ phase was obtained
 for $\lambda_1$  sufficiently positive.
 In the case of $SU(3)$, the interplay of potential
 terms when $\lambda_1$ is on the order of $T^3$
 leads to a skewed phase \cite{Myers:2007vc}.

\begin{figure}
\centering
\includegraphics[width=.5\textwidth]{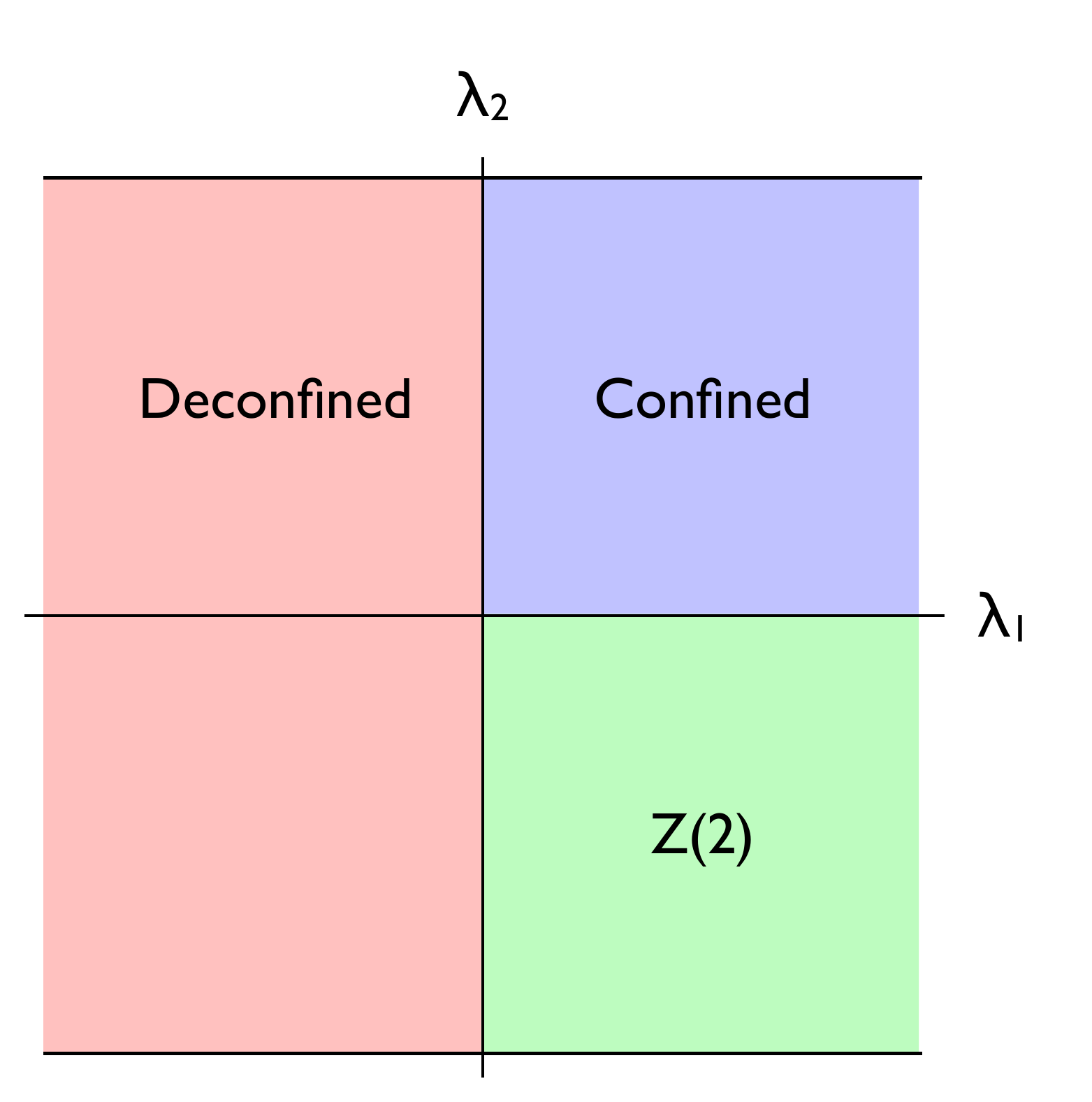}
\caption{Phase diagram for $SU(4)$.}
\label{fig:fig1}
\end{figure}

In the case of $SU(6)$, there are two non-trivial subroups
of $Z(6)$: $Z(2)$ and $Z(3)$.
In the phase where
there is an unbroken $Z(2)$ global symmetry,
we have
$\left< Tr_F P^2 \right> \neq 0$ and $\left< Tr_F P \right> = \left< Tr_F P^3 \right> = 0$.
If $Z(6)$ breaks to 
$Z(3)$, we have instead
$\left< Tr_F P^3 \right> \neq 0$ and $\left< Tr_F P \right> = \left< Tr_F P^2 \right> = 0$.
The phase structure for $SU(6)$, obtained by
minimizing $V_{ext}$
is shown in Fig. 2 for the case $\lambda_1 > 0$ and
in Fig. 3 for the case $\lambda_1 < 0$.
Note that the confined and deconfined phases are not directly connected
in the case $\lambda_1>0$, but are separated by the other phases.
As in the case of $SU(4)$, 
$O(T^3)$ corrections
shift the critical lines somewhat, but
preserve the overall structure of the phase diagram.

\begin{figure}[!h]
  \hfill
  \begin{minipage}[t!]{.45\textwidth}
    \begin{center}  
      \includegraphics[width=0.95\textwidth]{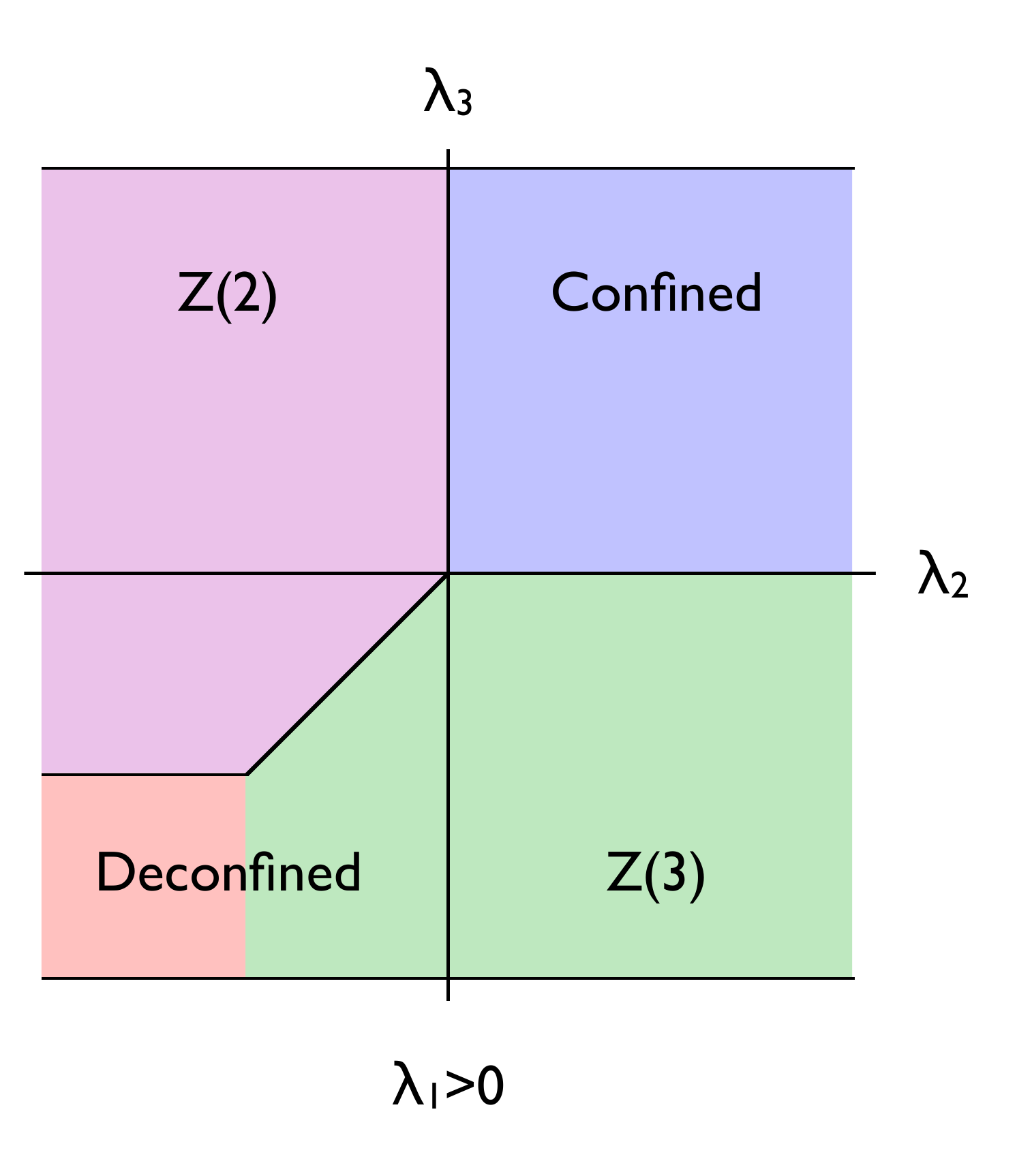}
      \vspace{-4mm}
      \caption{$SU(6)$ phases for $\lambda_1>0$.}
      \label{fig:2}
    \end{center}
  \end{minipage}
  \begin{minipage}[t!]{.45\textwidth}
    \begin{center}  
      \includegraphics[width=0.95\textwidth]{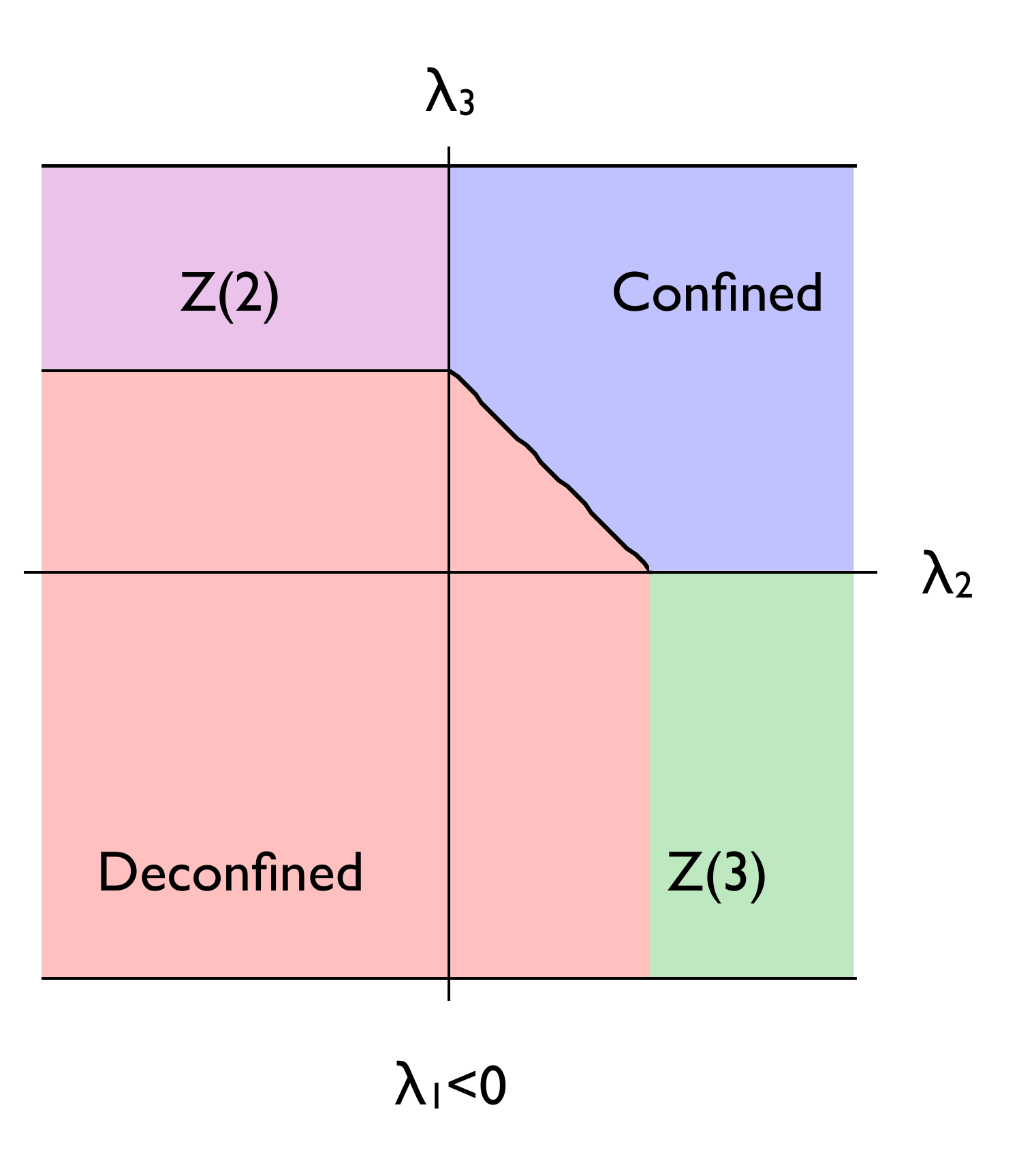}
      \vspace{-4mm}
      \caption{$SU(6)$ phases for $\lambda_1<0$.}
      \label{fig:3}
    \end{center}
  \end{minipage}
  \hfill
\end{figure}

%\begin{figure}
%\centering
%\includegraphics[width=.5\textwidth]{su6phases-lampos.pdf}
%\caption{Test pdf.}
%\label{fig:Test_pdf}
%\end{figure}

%\begin{figure}
%\centering
%\includegraphics[width=.5\textwidth]{su6phases-lamneg.pdf}
%\caption{Test pdf.}
%\label{fig:Test_pdf}
%\end{figure}

\section{Properties of Partially Confined Phase}

For an $SU(N)$ gauge theory, we can associate a phase with
every integer decomposition of the form
$N = L M$, where  $Z(L)$ is unbroken, and $Z(M)$ is broken.
The value of $P$ at the minimum of the effective potential
can be written as an element of
$SU(L) \otimes SU(M)$ of the form
$P = P_{0L} \otimes 1_M$,
where $P_{0L}$ is associated with
the confined phase of $SU(L)$.
This $P$ has the property that 
$Tr_{F}\left( P^{(0)}_L \right)^k=0$
for all $k=1..\left(L-1\right)$,
and also satisfies 
$( P^{(0)}_L )^L=1_L$.
The value of the effective potential
at the minimum determines the free density
and other thermodynamic properties.

String tensions and screening masses are obtained
from small fluctuations about the minimum of the
effective potential in the usual way.
For the sake of simplicity, we describe only
typical results here, and generally omit corrections
due to the 
gauge boson term in the effective potential.
The simplest cases are the confined and deconfined
phases. 
In the confined phase, the string tensions $\sigma_k$ are
 obtained from
\begin{equation}
\left< Tr P^k (\vec x ) Tr P^{\dagger k} (\vec y ) \right> \sim \exp \left[ - \frac{1}{T} \sigma_k | \vec x - \vec y | \right]
\end{equation}
are given by
\begin{equation}
\sigma_k = k \sqrt{{\frac{1}{2}}g^2 N \lambda_k T}.
\end{equation}
It is quite striking that each $\sigma_k$ can be
varied independently. Of course, in regions
where, {\it e.g.}, $\sigma_2 > 2 \sigma_1$, we expect
to see that the asymptotic $k=2$ string tension
is $2 \sigma_1$, 
due to interactions neglected in the quadratic approximation.
In the deconfined phase, the Debye screening mass
$m_D$
is given by
\begin{equation}
m_D^2={{g^2N}\over{T}}\sum_{k=1}^{\left[ {N\over 2} \right]} k^2\lambda_k
\end{equation}
for Polyakov loops of all $N$-alities.

Partially confined phases have both string tensions
and screening masses. In each $k$-sector, the
square of the mass is a linear combination of the
parameters $\lambda_k$. In confined sectors,
the string tension $\sigma_k$ is the mass divided by $T$.
For example,
in the case where the $Z(4)$ symmetry of $SU(4)$
is spontaneously broken to $Z(2)$,
there is a single string tension for odd values
of $k$, and a single screening mass for even values
of $k$.
The string tension associated with $Tr_F P$ is given by
\begin{equation}
\sigma_1 =\sqrt{g^2NT(\lambda_1-4\lambda_2)}
\end{equation}
and the screening mass associated with $Tr_F P^2$  is
\begin{equation}
m_{D} =2g\sqrt{-N\lambda_2/T}
\end{equation}

In a phase where $Z(N)$ is broken to $Z(L)$,
$Tr_F P^L$ plays a special role.
There are $M$ equivalent thermal
states associated with the broken $Z(M)$ symmetry.
Taking 
$P = P^0_L\, \otimes e^{2\pi i k/N}$,
where $k=0,..,(M-1)$,
we see that 
$Tr P^L = N \exp \left[ 2 \pi i k / M \right]$.
There are one-dimensional kink solutions
of the equations of motion derived
from the effective action
that interpolate between these states.
These solutions have a finite action per unit
area, which are the 't Hooft loop surface
tensions $\rho_k$.

The general form of such a kink solution is
$P(z) = P^0_L\, \otimes e^{i\psi (z)}$.
It is easy to calculate $\rho_k$ via BPS techniques if one term dominates
in the effective potential.
For example, in the case
where $-\lambda_L \gg T^3$
and no other $\lambda_k$ is negative or too large,
we have
\begin{equation}
\rho_{k}= 16 \left( kL \right) 
\left( N-kL \right)\sqrt{\frac{\left(-\lambda_L\right)T}{g^2 N}}.
\end{equation}
This result gives a modified Casimir scaling law:
\begin{equation}
\frac{\rho_{k}}{\rho_1}=
\frac{kL\left( N-kL \right)}{N-L}.
\end{equation}
In general, Casimir scaling will hold whenever
the relevant part of the effective potential can be written
as a sum of adjoint representation terms,
as is the case with our external potential $V_{ext}$.
In the case where the gauge boson contribution
to the effective potential
dominates in the kink solution, we have
\begin{equation}
\rho_{k}=\frac{4\pi^2 \left( kL \right) \left( N-kL \right) T^2}{3\sqrt{3 g^2 N}}.
\end{equation}
This requires $T^3 \gg  | \lambda_L | $,
as in our SU(4) simulations, where $\lambda_2=0$.
This result again obeys Casimir scaling,
and represents a generalization for the
formula for $\rho_k$ in the deconfined phase
 \cite{Giovannangeli:2001bh}.

\section{Conclusions}

A large number of phases can be created at high temperatures by adding Polyakov loop terms to the action, including many partially confined phases.
Phase diagrams can be easily predicted.
String tensions can be varied almost arbitrarily, and their values predicted.
't Hooft loop surface tensions can also be calculated, and obey Casimir scaling for a large class of potentials.

The case of $SU(6)$ is particularly interesting.
In addition to the confined and deconfined phases,
there are two distinct partially confined phases,
associated with the $Z(2)$ and $Z(3)$ subgroups
of $Z(6)$. This is the natural model to test the
analytic predictions we have made for
string tensions, screening masses, and 
other quantities characterizing the different phases.

The original motivation for studying
lattice gauge theories with additional
couplings to Polyakov loops
was to study the variation of string
tension scaling behavior.
The results of our analytical calculations
indicate a class of extended models
for which the string tensions can be varied
continuously. On the other hand,
the surface tensions
associated with 't Hooft loops,
representing the $Z(N)$ duals
of Wilson loops,
{\it i.e.} magnetic monopole loops,
are predicted to obey Casimir
scaling in the deconfined and partially
confined phases.
We are beginning work to confirm
these analytical results with lattice simulations.
We are also planning a study
of the behavior of calorons
in these models,
which should offer a direct view, both
analytically and from simulations,
of the role of calorons in confining
gauge theories.


\begin{thebibliography}{99}

%\cite{Myers:2007vc}
\bibitem{Myers:2007vc}
  J.~C.~Myers and M.~C.~Ogilvie,
  \emph{New Phases of SU(3) and SU(4) at Finite Temperature},
  arXiv:0707.1869 [hep-lat].
  %%CITATION = ARXIV:0707.1869;%%
  
  %\cite{Meisinger:2002kg}
\bibitem{Meisinger:2002kg}
  P.~N.~Meisinger, T.~R.~Miller and M.~C.~Ogilvie,
  \emph{A phenomenological treatment of chiral symmetry restoration and
  deconfinement},
  \emph{Nucl.\ Phys.\ Proc.\ Suppl.}\  {\bf 119} (2003) 511
  [arXiv:hep-lat/0208073].
  %%CITATION = NUPHZ,119,511;%%
  
  %\cite{Meisinger:2005ws}
\bibitem{Meisinger:2005ws}
  P.~N.~Meisinger and M.~C.~Ogilvie,
   \emph{Non-universality of string tension ratios and gluon confinement at finite
  temperature},
  \emph{PoS} {\bf LAT2005} (2006)  195.
  %%CITATION = POSCI,LAT2005,195;%%
  
  %\cite{Meisinger:2006hg}
\bibitem{Meisinger:2006hg}
  P.~N.~Meisinger and M.~C.~Ogilvie,
   \emph{String tension scaling in models of the confined phase},
  \emph{PoS} {\bf LAT2006} (2006) 072 
  [arXiv:hep-lat/0612002].
  %%CITATION = POSCI,LAT2006,072;%%
    
  %\cite{Davies:1999uw}
\bibitem{Davies:1999uw}
  N.~M.~Davies, T.~J.~Hollowood, V.~V.~Khoze and M.~P.~Mattis,
   \emph{Gluino condensate and magnetic monopoles in supersymmetric  gluodynamics},
  \emph{Nucl.\ Phys.}\  B {\bf 559} (1999) 072 
  [arXiv:hep-th/9905015].
  %%CITATION = NUPHA,B559,123;%%
  
  %\cite{Davies:2000nw}
\bibitem{Davies:2000nw}
  N.~M.~Davies, T.~J.~Hollowood and V.~V.~Khoze,
   \emph{Monopoles, affine algebras and the gluino condensate},
  \emph{J.\ Math.\ Phys.}\  {\bf 44} (2003) 3640 
  [arXiv:hep-th/0006011].
  %%CITATION = JMAPA,44,3640;%%
  
  %\cite{Diakonov:2004jn}
\bibitem{Diakonov:2004jn}
  D.~Diakonov, N.~Gromov, V.~Petrov and S.~Slizovskiy,
   \emph{Quantum weights of dyons and of instantons with non-trivial holonomy},
  \emph{Phys.\ Rev.}\  D {\bf 70} (2004) 036003 
  [arXiv:hep-th/0404042].
  %%CITATION = PHRVA,D70,036003;%%
  
    %\cite{Meisinger:2001fi}
\bibitem{Meisinger:2001fi}
  P.~N.~Meisinger and M.~C.~Ogilvie,
   \emph{Complete high temperature expansions for one-loop finite temperature
  effects},
  \emph{Phys.\ Rev.}\  D {\bf 65} (2002) 056013
  [arXiv:hep-ph/0108026].
  %%CITATION = PHRVA,D65,056013;%%
  
%\cite{Gross:1980br}
\bibitem{Gross:1980br}
  D.~J.~Gross, R.~D.~Pisarski and L.~G.~Yaffe,
   \emph{QCD And Instantons At Finite Temperature},
  \emph{Rev.\ Mod.\ Phys.}\  {\bf 53} (1981) 43.
  %%CITATION = RMPHA,53,43;%%
  
  %\cite{Weiss:1980rj}
\bibitem{Weiss:1980rj}
  N.~Weiss,
   \emph{The Effective Potential For The Order Parameter Of Gauge Theories At Finite Temperature},
  \emph{Phys.\ Rev.}\  D {\bf 24} (1981) 475 .
  %%CITATION = PHRVA,D24,475;%%  
  
 %\cite{Bhattacharya:1990hk}
\bibitem{Bhattacharya:1990hk}
  T.~Bhattacharya, A.~Gocksch, C.~Korthals Altes and R.~D.~Pisarski,
   \emph{Interface tension in an SU(N) gauge theory at high temperature},
  \emph{Phys.\ Rev.\ Lett.}\  {\bf 66} (1991) 998.
  %%CITATION = PRLTA,66,998;%%

%\cite{Bhattacharya:1992qb}
\bibitem{Bhattacharya:1992qb}
  T.~Bhattacharya, A.~Gocksch, C.~Korthals Altes and R.~D.~Pisarski,
   \emph{Z(N) interface tension in a hot SU(N) gauge theory},
  \emph{Nucl.\ Phys.}\  B {\bf 383} (1992) 497
  [arXiv:hep-ph/9205231].
  %%CITATION = NUPHA,B383,497;%%
  
  %\cite{Bursa:2005yv}
\bibitem{Bursa:2005yv}
  F.~Bursa and M.~Teper,
   \emph{Casimir scaling of domain wall tensions in the deconfined phase of D =  3+1 SU(N) gauge theories},
  \emph{JHEP} {\bf 0508} (2005) 060
  [arXiv:hep-lat/0505025].
  %%CITATION = JHEPA,0508,060;%%

%\cite{de Forcrand:2005rg}
\bibitem{de Forcrand:2005rg}
  P.~de Forcrand, B.~Lucini and D.~Noth,
   \emph{'t Hooft loops and perturbation theory},
  \emph{PoS} {\bf LAT2005} (2006) 323
  [arXiv:hep-lat/0510081].
  %%CITATION = POSCI,LAT2005,323;%%
  
%\cite{Giovannangeli:2001bh}
\bibitem{Giovannangeli:2001bh}
  P.~Giovannangeli and C.~P.~Korthals Altes,
   \emph{'t Hooft and Wilson loop ratios in the QCD plasma},
  \emph{Nucl.\ Phys.}\  B {\bf 608} (2001) 203
  [arXiv:hep-ph/0102022].
  %%CITATION = NUPHA,B608,203;%%
    
\end{thebibliography}
\end{document}